\documentclass{llncs}
\usepackage{amsmath, amsfonts, amssymb}
\usepackage{graphicx, xypic,xy}

\begin{document}

\title{A Logic for Strategy Updates}

\author{Can Ba\c{s}kent\inst{}}

\institute{Department of Computer Science, Graduate Center, City University of New York  \email{cbaskent@gc.cuny.edu ~ www.canbaskent.net}}

\maketitle


\section{Introduction}

In game theory, \emph{strategy} for a player is defined as ``a set of rules that describe exactly how (...) [a] player should choose, depending on how the [other] players have chosen at earlier moves" \cite{hod}. Notice that this definition of strategies is \emph{static}, and presumably constructed \emph{before} the game is actually played. 

For example, consider chess. According to Zermelo's well-known theorem, chess is determined \cite{sch}. Then, why would you play chess if you know you will lose (or won't win) the game? Clearly, if we have logical omniscience (which we don't), then it is pointless for the player, who is going to lose, to even start playing the game as she knows the outcome already. If we are not logical omniscient, and have only a limited amount of computational power and memory (which we do), then chess is only a perfect information game for God. Therefore, there seems to be a problem. The static, pre-determined notion of strategies falls short analyzing perfect information games. Because, we, people, do not strategize as such even in perfect information games - largely because we are not logically omniscient, and we have limited memory, and computational and deductive power. 

\begin{figure}[b]
\begin{center}\mbox{
\begin{xy}
(-80,0)="A";
(-80,-7)*{(1, 0)};
(-80,-3)*{s_{6}};
(-80,20)*+{\bullet}="B";
{\ar "B";"A"};
(-80,23)*{s_{0}, P1};
(-70,22)*{a};
(-82,10)*{d};
(-60,0)="C";
(-60,-7)*{(0, 2)};
(-60,-3)*{s_{7}};
(-60,20)*+{\bullet}="D";
{\ar "D";"C"};
(-60,23)*{s_{1}, P2};
(-50,22)*{a};
(-62,10)*{d};
{\ar "B";"D"};
(-40,0)="E";
(-40,-7)*{(3, 1)};
(-40,-3)*{s_{8}};
(-40,20)*+{\bullet}="F";
{\ar "F";"E"};
(-40,23)*{s_{2}, P1};
(-30,22)*{a};
(-42,10)*{d};
{\ar "D";"F"};
(-20,0)="G";
(-20,-7)*{(2, 4)};
(-20,-3)*{s_{9}};
(-20,20)*+{\bullet}="H";
{\ar "H";"G"};
(-20,23)*{s_{3}, P2};
(-22,10)*{d};
(-10,22)*{a};
{\ar "F";"H"};
(0,0)="I";
(0,-7)*{(5, 2)};
(0,-3)*{s_{10}};
(0,20)*+{\bullet}="J";
{\ar "J";"I"};
(0,23)*{s_{4}, P1};
(-2,10)*{d};
(10,22)*{a};
(20,20)="K";
{\ar "J";"K"};
(29,20)*{~ (4, 6)};
(23,20)*{s_{5}};
{\ar "H";"J"};  \end{xy}
}
\end{center}
\caption{\emph{Centipede game}}
\label{cent}\end{figure}
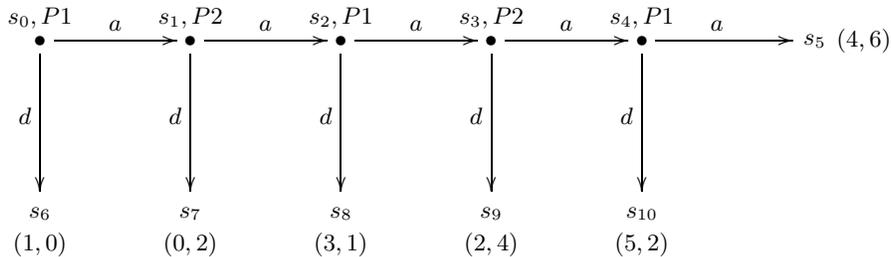

While people play games, they observe, learn, recollect and update their strategies \emph{during} the game as well as adopting deontological strategies and goals before the game. Players update and revise their strategies, for instance, when their opponent makes an \emph{unexpected} or \emph{irrational} move. Similarly, sometimes external factors may force the players not to make some certain moves. For instance, assume that you are playing a video game by using a gamepad or a keyboard, and in the middle of the game, one of the buttons on the gamepad brakes. Hence, from that moment on, you will not be able to make some moves in the game that are controlled by that button on the gamepad. This is most certainly not part of your strategy. Therefore, you will need to revise your strategy in such a way that some moves will be excluded from your strategy from that moment on. However, for your opponent, that is not the case as she can still make all the moves available to her.

Moreover, in some cases, assumptions about the game or the players may fail as well. For instance, consider the centipede game between two players $P1$ and $P2$ (Figure~\ref{cent}). Under the assumption of common rationality, the usual backward induction scheme produces the solution that $P1$ needs to make a $d$ move at $s_0$. What happens then, if $P1$ is prohibited or prevented from making $d$ move at $s_0$ and onwards right after the beginning of the game (or similarly, if the key on the gamepad that is used to make a $d$ move is broken)? It means that from a behavioral perspective, the assumption of common rationality is violated and $P2$ may need to update her strategy \emph{during the game} based on what she has observed. There can be many reasons why $P1$ may make such a move. The move $d$ might have been prohibited for $P1$ right after the game has started, or it may be a taboo for that specific player to make that move, or it may have been simply forbidden or restricted by an external factor (nature, God etc.). 

In this paper, we focus on what we call \emph{move updates} where some moves become unavailable during the game. Clearly, there can be considered many other forms of updates, revisions and restrictions that can happen during the game play. For instance, some states may become unavailable in the midst of the game for some players. Moreover, manipulation games where a third party or God/nature affects the outcome of the game are also examples of such games where dynamic strategy analysis is much needed.

Our goal here is to present a formal framework for move based strategy restrictions by extending \emph{strategy logic} (henceforth, SL) which was introduced by Ramanujam and Simon~\cite{ram0}. The rest of the paper is structured as follows. Section~\ref{SL} provides a short reminder of SL. In Section~\ref{DSL}, the framework of SL is extended with strategic move restrictions, and completeness of the resulting logic is shown. Then, we investigate some decision theoretical problems and give a complexity bound for the model checking problem in SL - which was an open problem so far. Finally, we conclude by placing RSL in the context of related work, followed by a conclusion and ideas for future research. The Appendix contains proofs of all propositions and theorems stated in Section~\ref{DSL}.

\section{Strategy logic}\label{SL}

In this section, we give a short overview of the strategy logic \cite{gho,ram0}. The focus is on games played between two players given by the set $N= \{ 1, 2 \}$, and a single set of moves $\Sigma$ for both. Let $\mathbf{T} = (S, \Rightarrow, s_0)$ be a tree rooted at $s_0 \in S$, on the set of vertices $S$. A partial function $\Rightarrow : S \times \Sigma \rightarrow S$ specifies the labeled edges of such a tree where labels represent the moves at the states. The extensive form game tree, then, is a pair $T = (\mathbf{T}, \lambda)$ where $\mathbf{T}$ is a tree as defined before, and $\lambda: S \rightarrow N$ specifies whose turn it is at each state. A strategy $\mu^i$ for a player $i \in N$ is a function $\mu^i : S^i \rightarrow \Sigma$ where $S^i = \{ s \in S : \lambda(s) = i \}$. For player $i$ and strategy $\mu^i$, the strategy tree $T_{\mu} = (S_{\mu}, \Rightarrow_{\mu}, s_0, \lambda_{\mu})$ is the least subtree of $T$ satisfying the following two natural conditions:
\begin{enumerate}
\item  $s_0 \in S_{\mu}$; 
\item For any $s \in S_{\mu}$, if $\lambda(s)=i$, then there exists a unique $s' \in S_{\mu}$ and action $a$ such that $s \stackrel{a}{\Rightarrow_{\mu}} s'$. Otherwise, if $\lambda(s) \neq i$, then for all $s'$ with $s\stackrel{a}{\Rightarrow}s'$ for some $a$, we have $s \stackrel{a}{\Rightarrow_{\mu}} s'$.
\end{enumerate} 
In other words, in the strategy tree, the root is included, and for the states that belong to the strategizing player, a unique move is assigned to the player, and for the other player, all possible moves are considered. Notice that, in SL, strategies return unique moves. Nevertheless, we would still have a tree even if the strategies are set-valued.

The most basic constructions in SL are strategy specifications. First, for a given countable set $X$, a set of basic formulas $BF(X)$ is defined as follows, for $a \in \Sigma$: 
$$BF(X) : = x \in X ~|~ \neg \varphi ~|~ \varphi \wedge \varphi ~|~ \langle a \rangle \varphi$$
Let $P^i$ be a countable set of atomic observables for player $i$, with $P = P^1 \cup P^2$. The syntax of strategy specifications is given as follows for $\varphi \in BF(P^i)$: 
$$Strat^i(P^i) : = [\varphi \rightarrow a]^i ~|~ \sigma_1 + \sigma_2 ~|~ \sigma_1 \cdot \sigma_2$$

The specification $[\varphi \rightarrow a]^i$ at player $i$'s position stands for ``play $a$ whenever $\varphi$ holds". The specification $\sigma_1 + \sigma_2$ means that the strategy of the player conforms to the specification  $\sigma_1$ or $\sigma_2$ and $\sigma_1 \cdot \sigma_2$ means that the strategy of the player conforms to the specifications $\sigma_1$ and $\sigma_2$. 

Let $M = (T, V)$ where $T = (S, \Rightarrow, s_0, \lambda)$ is an extensive form game tree as defined before, and $V : S \rightarrow 2^P$ is a valuation function for the set of propositional variables $P$.  The truth of a formula $\varphi\in BF(P)$ is given as usual for the propositional, Boolean and modal formulas. 

The notion ``strategy $\mu$ {\em conforms} to specification $\sigma$ for player $i$ at state $s$" (notation $\mu, s \models_i \sigma$) is defined as follows, where $\mathbf{out}_{\mu}(s)$ denotes the unique outgoing edge at $s$ with respect to $\mu$.
\begin{quote}\begin{tabular}{lll}
$\mu, s \models_i [\varphi \rightarrow a]^i$ & ~ iff ~ & $M, s \models \varphi$ implies $\mathbf{out}_{\mu}(s) = a$\\
$\mu, s \models_i \sigma_1 + \sigma_2$ & ~  iff ~ & $\mu, s \models_i \sigma_1$ or $\mu, s \models_i \sigma_2$ \\
$\mu, s \models_i \sigma_1 \cdot \sigma_2$ & ~ iff ~ & $\mu, s \models_i \sigma_1$ and $\mu, s \models_i \sigma_2$ \\
\end{tabular}\end{quote}

Now, based on the strategy specifications, the syntax of the strategy logic SL is given as follows: 
$$ p ~|~  \neg \varphi ~|~ \varphi_1 \wedge \varphi_2 ~|~ \langle a \rangle \varphi ~|~ (\sigma)_i : a ~|~ \sigma \rightsquigarrow_i \psi$$
for $p\in P$, $a \in \Sigma$, $\sigma \in Strat^i(P^i)$, and $\psi \in BF(P^i)$. We read $(\sigma)_i : a$ as ``at the current state the strategy specification $\sigma$ for player $i$ suggests that the move $a$ can be played". Subsequently, we read $\sigma \rightsquigarrow_i \psi$ as ``following strategy $\sigma$ player $i$ can ensure $\psi$". The other connectives and modalities are defined as usual. 

We now define the set of available moves at a state $s$ as $moves(s) : = \{ a \in \Sigma : \exists s' \in S \text{ with } s \stackrel{a}{\Rightarrow} s'\}$. Then, based on $moves$, we inductively construct the set of enabled moves at state $s$ in strategy $\sigma$ as follows.
\begin{itemize}
\item $[\psi \rightarrow a]^i(s) ~ = 
\begin{cases} 
\{ a \} & :  ~\lambda(s)=i; M, s \models \psi, a \in moves(s) \\
\emptyset & : ~\lambda(s)=i; M, s \models \psi, a \notin moves(s) \\
\Sigma & :  \text{ otherwise}
\end{cases}$
\item $(\sigma_1 + \sigma_2)(s) = \sigma_1(s) \cup \sigma_2(s)$
\item $(\sigma_1 \cdot \sigma_2)(s) ~ = \sigma_1(s) \cap \sigma_2(s)$
\end{itemize}
The truth definition for the strategy formulas are as follows: 
\begin{quote}\begin{tabular}{lll}
$M, s \models \langle a \rangle \varphi$& ~ iff ~& $\exists s$ such that $s \stackrel{a}{\Rightarrow} s'$ and $M, s' \models \varphi$\\
$M, s \models (\sigma)_i : a$ & ~ iff ~ & $a \in \sigma(s)$ \\
$M, s \models \sigma \rightsquigarrow_i \psi$ & ~ iff ~ & $\forall s'$ such that $s \Rightarrow^{*}_{\sigma} s'$ in $T_s | \sigma$, \\
& & we have $M, s' \models \psi \wedge (\mathbf{turn}_i \rightarrow \mathbf{enabled}_{\sigma})$ \\
\end{tabular}\end{quote}
where $\sigma(s)$ is as before, and $\Rightarrow^{*}_{\sigma}$ denotes the reflexive transitive closure of $\Rightarrow_{\sigma}$. Furthermore, $T_s$ is the tree that consists of the unique path from the root ($s_0$) to $s$ and the subtree rooted at $s$, and $T_s | \sigma$ is the least subtree of $T_s$ that contains a unique path from $s_0$ to $s$ and from $s$ onwards, for each player $i$ node, all the moves enabled by $\sigma$, and for each node of the opponent player, all possible moves . The proposition $\mathbf{turn}_i$ denotes that it is $i$'s turn to play. Finally, define $\mathbf{enabled}_{\sigma} = \bigvee_{a \in \Sigma} (\langle a \rangle \top \wedge (\sigma)_i : a)$. Now, we give the axioms of SL.
\begin{itemize}
\item All the substitutional instances of the tautologies of propositional calculus
\item $[a] (\varphi \rightarrow \psi) \rightarrow ([a] \varphi \rightarrow [a] \psi)$
\item $\langle a \rangle \varphi \rightarrow [a] \varphi$
\item $\langle a \rangle \top \rightarrow ([\psi \rightarrow a]^{i} : a)_{i}$ 
for all $a \in \Sigma$
\item $[\mathbf{turn}_i \wedge \psi \wedge ([\psi \rightarrow a]^i)_i : a)] \rightarrow \langle a \rangle \top$
\item $\mathbf{turn}_i \wedge ([\psi \rightarrow a]^i)_i : c \leftrightarrow \neg \psi$ for all $a \neq c$
\item $(\sigma + \sigma')_i: a \leftrightarrow (\sigma : a)_i \vee (\sigma': a)_i$
\item $(\sigma \cdot \sigma')_i : a \leftrightarrow (\sigma : a)_i \wedge (\sigma': a)_i$
\item $\sigma \rightsquigarrow_i \psi \rightarrow [\psi \wedge \mathbf{inv}^{\sigma}_i(a, \psi) \wedge \mathbf{inv}^{\sigma}_{-i}(\psi) \wedge \mathbf{enabled}_{\sigma}]$
\end{itemize}
Here, $\mathbf{inv}_{i}^{\sigma}(a, \psi)= (\mathbf{turn}_i \wedge (\sigma)_i : a)  \rightarrow [a] (\sigma \rightsquigarrow_i \psi)$ which expresses the fact that after an $a$ move by $i$ which conforms to $\sigma$, the statement $\sigma \rightsquigarrow_i \psi$ continues to hold, and $\mathbf{inv}^{\sigma}_{-i}(\psi)=\mathbf{turn}_i \rightarrow \odot (\sigma \rightsquigarrow_i \psi)$ states that after any move of $-i$, $\sigma \rightsquigarrow_i \psi$ continues to hold. Here, $\bigcirc \varphi \equiv \bigvee_{a \in \Sigma} \langle a \rangle \varphi$ and $\odot \varphi \equiv  \neg \bigcirc \neg \varphi$.

Now, we discuss the inference rules that SL employs: modus ponens and generalization for $[a]$ for each $a\in\Sigma$. The induction rule is a bit more complex: From the formulas $\varphi \wedge (\mathbf{turn}_i \wedge (\sigma)_{i} : a) \rightarrow [a] \varphi, \varphi \wedge \mathbf{turn}_{-i} \rightarrow \odot \varphi$, and $\varphi \rightarrow \psi \wedge \mathbf{enabled}_{\sigma}$ derive $\varphi \rightarrow \sigma \rightsquigarrow_i \psi$. The axiom system of SL is sound and complete with respect to the given semantics \cite{ram0}.

\section{Restricted strategy logic}\label{DSL}

\subsection{Basics}

Let us now extend SL to \emph{restricted strategy logic}, henceforth RSL, by allowing move restrictions during the game. Recall that our motivation can be illustrated with the example of a gamepad/keyboard which gets broken during the game play disallowing the player to make some certain moves from that moment on.

We denote the move restriction by $[\sigma ! a ]^{i}$ for a strategy specification $\sigma$, and move $a$ for player $i$. Informally, after the move restriction of $\sigma$ by $a$, player $i$ will not be able to make an $a$ move. We incorporate restrictions in RSL at the level of strategy specifications. In SL, recall that strategies are functions. Therefore, they only produce one move per state. However, our dynamic take in strategies cover more general cases where strategies can offer a \emph{set} of moves to the player. Thus, in RSL, we define strategy $\mu^i$ as $\mu^i : S^i \rightarrow 2^\Sigma$. By $\mathbf{outr}_{\mu^i}(s)$ we will denote the set of moves returned by $\mu^i$ at $s$. Then, the extended syntax of strategy specifications for player $i$ is given as follows.
$$ Strat^{i}(P^{i}): = [\psi \rightarrow a]^{i} ~|~ \sigma + \sigma ~|~ \sigma \cdot \sigma ~|~ [\sigma!a]^{i}$$
Notice that the restrictions affect only the player who gets a move restriction. In other words, if $a$ is prohibited to player $i$, it does not mean that some other player $j$ cannot make an $a$ move. In other words, if my gamepad/keyboard is broken, it doesn't mean that yours is broken as well. 

Once a move is restricted at a state, we will prone the strategy tree removing the prohibited move from that state on. Therefore, given $\mu^i : S^i \rightarrow 2^\Sigma$, we define the updated strategy relation $\mu^i!a: S^i \rightarrow 2^{\Sigma - \{ a\} }$. We are now ready to define confirmation of restricted specifications to strategies. Note that we skip the cases for $\cdot$ and $+$ as they are exactly the same.
\begin{quote}
\begin{tabular}{lll}
$\mu, s \models_i [\varphi \rightarrow a]^i$ & ~ iff ~ & $M, s \models \varphi$ implies $a \in \mathbf{outr}_{\mu}(s)$\\
$\mu^i, s \models_i [\sigma!a]^{i}$ & ~ iff ~ & $a \notin \mathbf{outr}_{\mu^i}(s)$ and $\mu!a, s \models_i \sigma$ 
\end{tabular}
\end{quote}
In the sequel, we omit the superscript that indicates the agents, thus, we  write $S$ for $S^{i}$, and $\sigma!a$ for $[\sigma ! a]^{i}$ when it is obvious. Given a strategy $\mu$ and its strategy tree $T_{\mu} = (S_{\mu}, \Rightarrow_{\mu}, s_{0}, \lambda_{\mu})$, we define the restricted strategy structure $T_{\mu!a}$ with respect to an action $a$. Once we removed the restricted moves, the updated structure may not be a tree (it may be a forest). For this reason, we take $(\mu!a,s)$ as the connected component of $T_{\mu!a}$ that includes $s$. Therefore, for a fixed strategy, restrictions may yield different restricted strategy trees at different states. This is perfectly fine for our intuition, because the state of the game where the restriction is made is important. In other words, it matters where may gamepad is broken during the game play. Now, for player $i$ and strategy $\mu^i$ and move $a$, the restricted strategy tree $T_{\mu!a} = (S_{\mu!a}, \Rightarrow_{\mu!a}, s', \lambda_{\mu!a})$ is the least subtree of $T$ satisfying the following two conditions:
\begin{enumerate}
\item  $s' \in S_{\mu!a}$; 
\item For any $s \in S_{\mu!a}$, if $\lambda(s)=i$, then for each action $b \neq a \in \mu!a(s)$, there exists a unique $t \in S_{\mu!a}$ such that $s \stackrel{b}{\Rightarrow_{\mu!a}} t$. Otherwise, if $\lambda(s) \neq i$, then for all $t$ with $s\stackrel{b}{\Rightarrow}t$ for $b \neq a$, we have $s \stackrel{b}{\Rightarrow_{\mu!a}} t$.
\end{enumerate} 

Note that we introduce such conditions so that an RSL model can easily be considered as a submodel of a SL model with some additional assumptions. We can now make some further observations. By the abuse of the notation, we will use $\leftrightarrow$ to denote the equivalence of strategy specifications with respect to the conformation relation.

\begin{proposition}
For any strategy $\mu$, state $s$, specification $\sigma$, and formula $\psi$, $\mu, s \not\models [[\psi \rightarrow a] ! a]$.
\end{proposition}

\begin{proposition}{\label{distribution}}
For strategy specifications $\sigma$ and $\sigma'$, and move $a$, $(\sigma \cdot \sigma') ! a \leftrightarrow (\sigma ! a) \cdot (\sigma' ! a)$ and $(\sigma + \sigma') ! a \leftrightarrow (\sigma ! a) + (\sigma' ! a)$.
\end{proposition} 

Restrictions stabilize immediately. For $n\geq 1$, we use notation $\sigma!^{n}a$ to denote $n \geq 1$-consecutive restrictions of $\sigma$ by move $a$. Similarly, we put $\mu!^{n}a$ for the corresponding strategy tree $\mu$. 

\begin{proposition}{\label{combined}}
For arbitrary strategy specification $\sigma$, move $a$ and state $s$, \\ $(\sigma ! a)!a \leftrightarrow \sigma ! a$. Moreover, we have $\sigma!^{n}a \leftrightarrow \sigma ! a$.
\end{proposition}

Since restrictions are local  and operate by elimination, the order of the restrictions does not matter.
\begin{proposition}{\label{mixed}}
For any moves $a$ and $b$, $(\sigma!a)!b \leftrightarrow (\sigma!b)!a$.
\end{proposition}

\subsection{A Case Study: The Centipede Game} 

Let us consider the centipede game (see Figure~\ref{cent}), and see how RSL can formalize it when a restricted strategy specification can change the game after an unexpected (or even irrational) move. Let us call the players $P1$ and $P2$. The set of actions in the centipede game is $\Sigma = \{ d, a \}$ where $d, a$ mean that the player moves down or across, respectively. Utilities for individual players are indicated by a tuple $(x, y)$ where $x$ is the utility for $P1$, and $y$ is the utility for $P2$. For the sake of generality, we will not impose any further conditions on the strategies.

In a recent work, Artemov approached the centipede game from a rationality and epistemology based point of view \cite{art7}. Now, similar to his approach, we will use symbols $\mathbf{r}_1$ and $\mathbf{r}_2$ to denote the propositions ``$P1$ is rational" and  ``$P2$ is rational", respectively. Let us  now construct rational strategies $\mu$ and $\nu$ for $P1$ and $P2$ respectively following the backward induction scheme. At $s_4$, $P1$ makes a $d$ move, if she is rational. Therefore, we have $\mu, s_4 \models [\mathbf{r}_1 \rightarrow d]^{P1}$. However, at $s_3$, $P2$ would be aware of $P1$'s possible move at $s_4$ and the fact that $P1$ is rational as well, thus makes a $d$ move if she herself is rational. So, we have $\nu, s_3 \models [(\mathbf{r}_2 \wedge (\langle a \rangle \mathbf{r}_1 \vee \langle d \rangle \mathbf{r}_1)) \rightarrow d]^{P2}$. Following the same strategy, we obtain the following.

$\mu, s_2 \models [\mathbf{r}_1 \wedge (\langle a \rangle (\mathbf{r}_2 \wedge (\langle a \rangle \mathbf{r}_1 \vee \langle d \rangle \mathbf{r}_1) \vee \langle d \rangle (\mathbf{r}_2 \wedge (\langle a \rangle \mathbf{r}_1 \vee \langle d \rangle \mathbf{r}_1))) \rightarrow d]^{P1}$

$\nu, s_1 \models [ \mathbf{r}_2 \wedge (\langle a \rangle (\mathbf{r}_1 \wedge (\langle a \rangle (\mathbf{r}_2 \wedge (\langle a \rangle \mathbf{r}_1 \vee \langle d \rangle \mathbf{r}_1) \vee \langle d \rangle (\mathbf{r}_2 \wedge (\langle a \rangle \mathbf{r}_1 \vee \langle d \rangle \mathbf{r}_1)))) \langle d \rangle (\mathbf{r}_1 \wedge (\langle a \rangle (\mathbf{r}_2 \wedge (\langle a \rangle \mathbf{r}_1 \vee \langle d \rangle \mathbf{r}_1) \vee \langle d \rangle (\mathbf{r}_2 \wedge (\langle a \rangle \mathbf{r}_1 \vee \langle d \rangle \mathbf{r}_1)))) )  \rightarrow d]^{P2}$

$\mu, s_0 \models [(\mathbf{r}_1 \wedge (\langle a \rangle (\mathbf{r}_2 \wedge (\langle a \rangle (\mathbf{r}_1 \wedge (\langle a \rangle (\mathbf{r}_2 \wedge (\langle a \rangle \mathbf{r}_1 \vee \langle d \rangle \mathbf{r}_1) \vee \langle d \rangle (\mathbf{r}_2 \wedge (\langle a \rangle \mathbf{r}_1 \vee \langle d \rangle \mathbf{r}_1)))) \langle d \rangle (\mathbf{r}_1 \wedge (\langle a \rangle (\mathbf{r}_2 \wedge (\langle a \rangle \mathbf{r}_1 \vee \langle d \rangle \mathbf{r}_1) \vee \langle d \rangle (\mathbf{r}_2 \wedge (\langle a \rangle \mathbf{r}_1 \vee \langle d \rangle \mathbf{r}_1)))) )) \vee \langle d \rangle(\mathbf{r}_2 \wedge (\langle a \rangle (\mathbf{r}_1 \wedge (\langle a \rangle (\mathbf{r}_2 \wedge (\langle a \rangle \mathbf{r}_1 \vee \langle d \rangle \mathbf{r}_1) \vee \langle d \rangle (\mathbf{r}_2 \wedge (\langle a \rangle \mathbf{r}_1 \vee \langle d \rangle \mathbf{r}_1)))) \langle d \rangle (\mathbf{r}_1 \wedge (\langle a \rangle (\mathbf{r}_2 \wedge (\langle a \rangle \mathbf{r}_1 \vee \langle d \rangle \mathbf{r}_1) \vee \langle d \rangle (\mathbf{r}_2 \wedge (\langle a \rangle \mathbf{r}_1 \vee \langle d \rangle \mathbf{r}_1)))) ))) \rightarrow d]^{P1}$

Let $\Diamond \varphi := \langle a \rangle \varphi \vee \langle d \rangle \varphi$. Then, we have the following statements.

$\mu, s_4 \models [\mathbf{r}_1 \rightarrow d]^{P1}$
 
$\nu, s_3 \models [\mathbf{r}_2 \wedge \Diamond \mathbf{r}_1 \rightarrow d]^{P2}$

$\mu, s_2 \models [\mathbf{r}_1 \wedge \Diamond (\mathbf{r}_2 \wedge \Diamond \mathbf{r}_1) \rightarrow d]^{P1}$

$\nu, s_1 \models [\mathbf{r}_2 \wedge \Diamond(\mathbf{r}_1 \wedge \Diamond (\mathbf{r}_2 \wedge \Diamond \mathbf{r}_1)) \rightarrow d]^{P2}$

$\mu, s_0 \models [ \mathbf{r}_1 \wedge \Diamond (\mathbf{r}_2 \wedge \Diamond(\mathbf{r}_1 \wedge \Diamond (\mathbf{r}_2 \wedge \Diamond \mathbf{r}_1))) \rightarrow d]^{P1}$

Therefore, backward inductively, under the assumption of common rationality, we observe that $P1$ should make a $d$ move at $s_0$. Furthermore, this can be generalized to many other games.

\begin{theorem} 
Assuming common rationality in RSL framework, backward induction scheme produces a unique solution in games with ordinal pay-offs.
\end{theorem}

Argument for the proof of the theorem is quite straight-forward. Even if the strategies may be set valued, and hence return a set of moves per state, the assumption of common rationality forces the player to choose the move which returns the highest pay-off. Since this fact is known among players, by induction, we can show that the solution is unique.

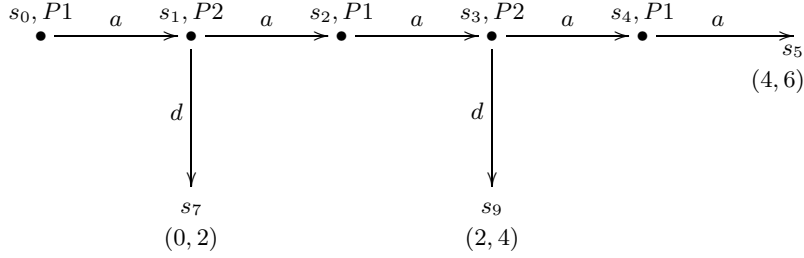
\begin{figure}[t]{\label{mua}}
\centering
\begin{xy}
(-80,20)*+{\bullet}="B";
(-80,23)*{s_{0}, P1};
(-70,22)*{a};
(-60,0)="C";
(-60,-7)*{(0, 2)};
(-60,-3)*{s_{7}};
(-60,20)*+{\bullet}="D";
{\ar "D";"C"};
(-60,23)*{s_{1}, P2};
(-50,22)*{a};
(-62,10)*{d};
{\ar "B";"D"};
(-40,20)*+{\bullet}="F";
(-40,23)*{s_{2}, P1};
(-30,22)*{a};
{\ar "D";"F"};
(-20,0)="G";
(-20,-7)*{(2, 4)};
(-20,-3)*{s_{9}};
(-20,20)*+{\bullet}="H";
{\ar "H";"G"};
(-20,23)*{s_{3}, P2};
(-22,10)*{d};
(-10,22)*{a};
{\ar "F";"H"};
(0,20)*+{\bullet}="J";
(0,23)*{s_{4}, P1};
(10,22)*{a};
(20,20)="K";
{\ar "J";"K"};
(18,14)*{(4, 6)};
(20,18)*{s_{5}};
{\ar "H";"J"};  
\end{xy}\label{mua}
\caption{Restricted centipede game-tree}
\end{figure}

Let us follow the backward induction scheme again with the updated game tree given above.  Notice that the specification $[\mathbf{r}_1 \rightarrow d]^{P1}$ does not conform with $\mu!d$ (as $\mu!d, s \not\models [[\mathbf{r}_1 \rightarrow d]^{P1}!d]^{P1}$ for all $s$)\footnote{We slightly abuse the formal language here. For a specification $\sigma$, we put $\mu, s \models \neg \sigma$ if it is not the case that $\mu, s \models \sigma$.}. Therefore, after the move restriction rational move for $P1$ is not admissible for her from that point on. Thus, $\mu!d$ conforms to those specifications that implies an $a$ move. Thus, $\mu!d, s_4 \models [\top \rightarrow a]^{P1} \cdot \neg [\mathbf{r}_1 \rightarrow d]^{P1}$. Therefore, at $s_3$, being rational, $P2$ choses the move with the highest pay-off, and makes an $a$ move. Thus, $\nu, s_3 \models \mathbf{r}_2 \wedge (\Diamond \neg [\mathbf{r}_1 \rightarrow d]) \rightarrow a]^{P2}$. Similarly, at $s_1$, we have $\nu, s_1 \models \mathbf{r}_2 \wedge (\Diamond \neg [\mathbf{r}_1 \rightarrow d] \wedge \Diamond (\mathbf{r}_2 \wedge (\Diamond \neg [\mathbf{r}_1 \rightarrow d]))) \rightarrow a]^{P2}$. Finally, at the root, we have $\mu!d, s_0 \models [\top \rightarrow a]^{P1}$. Thus, in the restricted centipede game, clearly, $P1$ has to make an $a$ move.

Even if it is not in the scope of this paper, this example also shows that we can view rationality as a \emph{set of restrictions}. Namely, some restrictions forces the players to play irrationally, while some restrictions allow them to play rationally. RSL, in this respect, gives a framework where both rational and irrational strategies can be analyzed.

\subsection{Axiomatization, Completeness, and Complexity}{\label{RSL-axiom}}

Before discussing the axiomatization of RSL, we note that the set of available moves $moves$ is defined as previously. Then, we define the set of enabled moves for the move restriction operator as $([\sigma ! a]^{i})(s) = \sigma(s) - \{ a \}$. Namely, if the move $a$ is not allowed any more, it should not be enabled. 

We now give the syntax of RSL, which is the same as that of SL.
$$ p ~|~ (\sigma)_i : a ~|~ \neg \varphi ~|~ \varphi_1 \wedge \varphi_2 ~|~ \langle a \rangle \varphi ~|~ \sigma \rightsquigarrow_i \psi$$

The semantics and the truth definitions of the formulas are defined as earlier with the exception of strategy specifications for restrictions (cf. Section 2). The axiom system of RSL consists of the axioms and rules of SL together with the following additional axiom for the added specification construct.
\begin{itemize} \item $(\sigma!a)_i : c \leftrightarrow \mathbf{turn}_i \wedge \neg ((\sigma)_i : a) \wedge (\sigma)_i : c$\end{itemize}
The soundness of the given axiom is straightforward and hence skipped. The new specification we introduced does not bring along an extra derivation rule since the new operator is at the level of specifications, not the formulas. 

Now, we observe that restricted moves are not enabled. We state it as a theorem with an immediate proof.

\begin{theorem}
$(\sigma!a)_i : a \leftrightarrow \bot$.
\end{theorem}

As expected, the system RSL is complete.

\begin{theorem}{\label{bgv-completeness}}
RSL is complete with respect to the given semantics.
\end{theorem}

However, the decidability of RSL and SL are not immediate. To the best of our knowledge, SL has not yet been proved to be decidable (or undecidable). Here, we show an upper bound for the model checking problem for both logics.

Now, we discuss a reduction of SL to (multi-modal) Computational Tree Logic CTL*. We refer the readers who are not familiar with CTL* to \cite{eme,eme0}. First, we discuss how to construct a CTL* model based on a given SL/RSL model, then describe a translation from SL to CTL*. 

Let us take a strategy model $M = (T, V)$ where $T = (S, \Rightarrow, s_0, \lambda)$. Notice that the function $\Rightarrow$ was defined from $S \times \Sigma$ to $S$. We can \emph{redefine} it by \emph{currying}. Given $\Rightarrow : S \times \Sigma \rightarrow S$, we can define the transition $\stackrel{a}{\Rightarrow} : S \rightarrow S$ for each move $a$. Therefore, we can curry $\Rightarrow $ to get $\Rightarrow = \bigcup_{a \in \Sigma} \stackrel{a} {\Rightarrow}$. We can think of $M$ as a pointed multi-modal CTL* tree model $M^{*} = (S, s_0, \{ \stackrel{a}{\Rightarrow} \}_{a \in \Sigma}, V)$ that has next time modalities for each action. In this case, corresponding to each binary relation $\stackrel{a}{\Rightarrow}$, we will have a dynamic next-time modality $\mathsf{X}_a$ which quantifies over the sub-path on the same branch (note that state formulas are also path formulas \cite{eme0}). For the reflexive-transitive closure of $\stackrel{a}{\Rightarrow}$, we will use $\Box$ for all accessible future times in the same branch. Before giving the translation of SL specifications and formulas into CTL*, let us introduce some special propositions. We label the states that are returned by a strategy $\mu$ with the proposition $\mathbf{strategy}_{\mu}$, stipulating that $\mathbf{strategy}_{\mu}$ holds only at those points. Notice that given two points in the domain of $\mu$, there is a unique path between these two (which satisfies  $\mathbf{strategy}_{\mu}$). Moreover, we use $\mathbf{turn}_i$ as a proposition that denotes that it is $i$'s turn to play, i.e. $s \models \mathbf{turn}_{i}$ iff $s \in S^{i}$. 

We now give two translations. First, $\mathbf{tr}$ translates strategy specifications to CTL* formulas while the second $\mathbf{Tr}$ translates SL formulas to CTL* formulas. Given a strategy $\mu$, conformation to $\mu$ is translated as follows.

\begin{itemize}
\item $\mathbf{tr}([\psi \rightarrow a]^{i}) = \mathbf{Tr}(\psi)
\rightarrow \mathsf{E}(\mathbf{strategy}_{\mu} \wedge \mathsf{X}_{a}
\top)$
\item $\mathbf{tr}(\sigma_1 + \sigma_2) = \mathbf{tr}(\sigma_{1}) \vee
\mathbf{tr}(\sigma_{2})$
\item $\mathbf{tr}(\sigma_1 \cdot \sigma_2) = \mathbf{tr}(\sigma_{1})
\wedge \mathbf{tr}(\sigma_{2})$
\end{itemize}

Notice that in the first case, the translation makes use of the formula translation $\mathbf{Tr}$ for formula $\psi$ as defined below. Assuming the correctness of $\mathbf{Tr}$ (which we show in Theorem~\ref{bgv-translation}), correctness of the translation $\mathbf{tr}$ is straightforward.

\begin{theorem}{\label{bgv-spec-translation}}
Let $\mu$ be a strategy and $\sigma$ a strategy specification in SL, and let $\mu^{*}$ be the corresponding (sub)tree in a CTL* model. Then, $\mu, s \models \sigma$ iff $\mu^{*}, s \models \mathbf{tr} (\sigma)$.
\end{theorem}

Here follows the translation $\mathbf{Tr}$ of formulas from SL to CTL* skipping the Boolean cases. Note that the translation is very similar to the Kripke semantics for Propositional Dynamic Logic where for each action $a$, a relation $R_a$ and a modality $\langle a \rangle$ associated with $R_a$ are introduced.

\begin{itemize}
\item $\mathbf{Tr} (\langle a \rangle \varphi) = \mathsf{X}_{a}
\mathbf{Tr}(\varphi)$
\item $\mathbf{Tr}((\sigma)_i : c) = \mathsf{X}_{c} \top$ for $c \in
\sigma(s)$
\item $\mathbf{Tr}((\sigma)_i : c) = \bot$ if $c \notin \sigma(s)$
\item $\mathbf{Tr}(\sigma \rightsquigarrow_i \psi) = \mathsf{E} \Box
(\mathbf{strategy}_{\sigma} \wedge (\mathbf{Tr}(\psi) \wedge
(\mathbf{turn}_i \rightarrow
\mathbf{enabled}_{\sigma})))$
\end{itemize}

Now, the atom $\mathbf{enabled}_{\sigma}$ is true at a state $s$ in CTL* if and only if for at least one $a \in \Sigma$ we have $\mathsf{X}_a \top$ and $a \in \sigma(s)$. Finally, the CTL* correspondence of $\sigma(s)$ to the set of enabled moves at $s$ by the strategy specification $\sigma$ is defined exactly as before with one small arrangement in the definition of admissible moves at a given state $s$, namely $moves(s) = \{ a : s \models_{CTL*} \mathsf{X}_a \top \}$. As an example, consider the translation of the proposition $\mathbf{out}_{\mu} = a$. Recall that it means that $a$ is the unique outgoing edge according to strategy $\mu$ at the state where the formula is interpreted. Therefore, there is a branch that is followed by strategy $\mu$, and at that branch, at the current state, $a$ is an admissible move. This corresponds to the translation $\mathsf{E} (\mathbf{strategy}_{\mu} \wedge \mathsf{X}_{a} \top)$. The following theorem summarizes our efforts here. 

\begin{theorem}{\label{bgv-translation}}
Let $M$ be a SL model and let $M^{*}$ be its CTL* correspondent. Then, $M, s \models \varphi$ iff $M^{*}, s \models \mathbf{Tr}(\varphi)$ for any state $s \in S$.
\end{theorem}

Note that the translation we suggest is model-dependent. For example, the predicate $\mathbf{strategy}_{\mu}$, depends on the strategy $\mu$, thus the model. For this reason, the suggested translation is not entirely syntactic and does not give us an immediate decidability result (using the fact that CTL* is decidable).
However, model checking problem for CTL* is PSPACE-complete \cite{eme}. Therefore, we have an upper bound for the complexity of the model checking problem for SL. We next observe that the model checking problem for both SL and RSL are in PSPACE.

\begin{theorem}{\label{bgv-decide}}
The model checking problem for SL is in PSPACE.
\end{theorem}
\begin{corollary}{\label{bgv-exponential}}
The model checking problem for RSL is in PSPACE.
\end{corollary}

\section{Related Work and Conclusion}\label{RL}

From a logical point of view, the idea of treating strategies as the ``unsung  heroes of games" can be traced back to van Benthem \cite{ben5,ben17}. In line with this program, Ramanujam, Simon, and Paul \cite{pau,ram,ram0,sim} have initiated a study within game theory and dynamic logic by taking strategies as the focus of concern and treating them as the primitives of games. They discuss strategies as a way of reasoning within and about games, and investigate how games and strategies behave under some further assumptions. 

The game-theoretical approach to the centipede game was initiated by Rosenthal \cite{ros}, which was followed by several solution methods \cite{aum0,hal3}. More recently, the role of rationality in its solution methods has been discussed \cite{bal0,ben14} and extended to some other cases \cite{pic}.The role knowledge as opposed to beliefs has been discussed in similar contexts by Artemov \cite{art7}. Moreover, based on experimental work, Ghosh, Meijering and Verbrugge aim to observe how humans reason strategically \cite{GMV10}. They consider simple centipede-like games from the cognitive point of view, and use strategy logic to formalize their findings.

This work is built on the aforementioned studies, and introduces a dynamic twist in formalizing strategies in games. What we suggest is a formal framework where restrictions in strategies are allowed during the game. We propose a new logic, called restricted strategy logic (RSL), to express strategy restrictions. We show its completeness and discuss model checking issues. Along the way, we also discuss model checking of SL, which has not been done so far (to the best of our knowledge). We claim that RSL presents a succinct way to represent strategy revisions. One direction for future research concerns the question on how RSL can be connected to the switching strategy frameworks. Additionally, one could come up with an restriction methodology where players change their rationality assumption. Note that, within the framework of RSL, different types of rationality (utilitiarian, deontological etc) can be seen as a different set of \emph{restrictions}. For instance, a player can initially commit herself to a $\max \min$ type of strategy and then change her commitment to a $\max \max$ type of strategy. Therefore, such changes can be represented from switching from restriction set $A$ to restriction set $A'$, for instance. In short, different \emph{types} of rationality can cause different revisions in players' strategies. We leave such analysis to future work.

Moreover, note that the way strategy restrictions work, and its completeness proof resemble public announcement logic \cite{dit,pla}. In the case of RSL, the move restriction can be seen as a prohibitive \emph{negative} announcement, corresponding to an elimination of moves that agree with the prohibition.

In conclusion, we believe that RSL presents a concise and natural framework for dynamic strategizing in games, and it can be extended in several thought-provoking ways.

\paragraph{Acknowledgements} We especially acknowledge the help and encouragement of Sujata Ghosh and Rineke Verbrugge. Sujata spent hours reading the paper and providing me with feedback, and Rineke read the paper very carefully many times and suggested lots of corrections. The idea of using CTL* for the decidability result and the name ``restricted strategy logic" have been suggested by Sujata. This paper is the product of author's visit to the Department of Artificial Intelligence of the University of Groningen. We also thank Sergei Artemov and Rohit Parikh for their feedback.

\bibliographystyle{plain} \bibliography{strategy}

\begin{thebibliography}{10}

\bibitem{art7}
Sergei Artemov.
\newblock Rational decisions in non-probablistic setting.
\newblock Technical Report TR-2009012, Department of Computer Science, The
  Graduate Center, The City University of New York, 2009.

\bibitem{aum0}
R.~J. Aumann.
\newblock Backward induction and common knowledge of rationality.
\newblock {\em Games and Economic Behavior}, 8(1):6--19, 1995.

\bibitem{bal0}
A.~Baltag, S.~Smets, and J.~Zvesper.
\newblock Keep `hoping' for rationality: A solution to the backward induction
  paradox.
\newblock {\em Synthese}, 169(2):301--333, 2009.

\bibitem{ben5}
J.~van Benthem.
\newblock Extensive games as process models.
\newblock {\em Journal of Logic, Language and Information}, 11:289--313, 2002.

\bibitem{ben17}
J.~van Benthem.
\newblock In praise of strategies.
\newblock In J.~{van Eijck} and R.~Verbrugge, editors, {\em Games, Actions and
  Social Software}, Texts in Logics and Games. Springer Verlag, Berlin, 2011.

\bibitem{ben14}
J.~van Benthem and A.~Gheerbrant.
\newblock Game solution, epistemic dynamics and fixed-point logics.
\newblock {\em Fundamenta Informaticae}, 100:19--41, January 2010.

\bibitem{bla1}
Patrick Blackburn, Maartijn~de Rijke, and Yde Venema.
\newblock {\em Modal Logic}.
\newblock Cambridge Tracts in Theoretical Computer Science. Cambridge
  University Press, 2001.

\bibitem{dit}
H.~van Ditmarsch, W.~{van der Hoek}, and B.~Kooi.
\newblock {\em Dynamic Epistemic Logic}.
\newblock Springer, Berlin, 2007.

\bibitem{eme}
A.E. Emerson.
\newblock Temporal and modal logic.
\newblock In J.~van Leeuwen, editor, {\em Handbook of Theoretical Computer
  Science}, volume~B, pages 995--1072. Elsevier and MIT-Press, Amsterdam and
  Cambridge (MA), 1990.

\bibitem{eme0}
E.A. Emerson and J.Y. Halpern.
\newblock {S}ometimes' and `not never' revisited: On branching versus linear
  time temporal logic.
\newblock {\em Journal of the Association for Computing Machinery},
  33:151--178, 1986.

\bibitem{GMV10}
S.~Ghosh, B.~Meijering, and R.~Verbrugge.
\newblock Logic meets cognition: Empirical reasoning in games.
\newblock In O.~Boissier et~al., editor, {\em MALLOW}, volume 627 of {\em CEUR
  Workshop Proceedings}. CEUR-WS.org, 2010.

\bibitem{gho}
S.~Ghosh, R.~Ramanujam, and S.~Simon.
\newblock On strategy composition and game composition.
\newblock Unpublished manuscript, 2010.

\bibitem{hal3}
J.~Y. Halpern.
\newblock Substantive rationality and backward induction.
\newblock {\em Games and Economic Behavior}, 37(2):425--435, November 2001.

\bibitem{hod}
Wilfred Hodges.
\newblock Logic and games.
\newblock In Edward~N. Zalta, editor, {\em The Stanford Encyclopedia of
  Philosophy}. 2009.

\bibitem{pau}
S.~Paul, R.~Ramanujam, and S.~Simon.
\newblock Dynamic restriction of choices: a preliminary logical report.
\newblock In A.~Heifetz, editor, {\em TARK}, pages 218--226, 2009.

\bibitem{pic}
M.~Piccione and A.~Rubinstein.
\newblock On the interpretation of decision problems with imperfect recall.
\newblock In Y.~Shoham, editor, {\em TARK}, pages 75--76. Morgan Kaufmann,
  1996.

\bibitem{pla}
J.~A. Plaza.
\newblock Logics of public communication.
\newblock In M.~L. Emrich, M.~S. Pfeifer, M.~Hadzikadic, and Z.~W. Ras,
  editors, {\em Proceedings 4th International Symposium on Methodologies for
  Intelligent Systems}, pages 201--216. Oak Ridge National Laboratory,
  ORNL/DSRD-24, 1989.

\bibitem{ram}
R.~Ramanujam and S.~Simon.
\newblock Dynamic logic on games with structured strategies.
\newblock In {\em Proceedings of the 11th International Conference on
  Principles of Knowledge Representation and Reasoning (KR-08)}, pages 49--58.
  AAAI Press, 2008.

\bibitem{ram0}
R.~Ramanujam and S.~Simon.
\newblock A logical structure for strategies.
\newblock In {\em Logic and the Foundations of Game and Decision Theory (LOFT
  7)}, volume~3 of {\em Texts in Logic and Games}, pages 183--208. Amsterdam
  University Press, 2008.

\bibitem{sim}
R.~Ramanujam and S.~Simon.
\newblock Reasoning in games.
\newblock In M.~K. Chakraborty, B.~L{\"{o}}we, and M.~N. Mitra, editors, {\em
  Logic, Navya-Nyaya and its Applications: Homage to Bimal Krishna
  Chakraborty}. College Publications, London, 2008.

\bibitem{ros}
R.~W. Rosenthal.
\newblock Games of perfect information, predatory pricing and the chain-store
  paradox.
\newblock {\em Journal of Economic Theory}, 25(1):92--100, 1981.

\bibitem{sch}
Ulrich Schwalbe and Paul Walker.
\newblock Zermelo and the early history of game theory.
\newblock {\em Games and Economic Behavior}, 34(1):123--137, January 2001.

\end{thebibliography}

\appendix
\section*{Appendix: Proofs}

\begin{proof} [Proposition \ref{distribution}]
Consider the first case. \vspace{-0.2cm}
\begin{quote} \begin{tabular}{lll}
$\mu, s \models (\sigma \cdot \sigma')!a$ & ~ iff ~ & $a \notin \mathbf{outr}_{\mu}(s)$ and $\mu!a, s \models \sigma \cdot \sigma'$ \\
& ~ iff ~ & $a \notin \mathbf{outr}_{\mu}(s)$ and $\mu!a, s \models \sigma$ and $\mu!a, s \models \sigma'$ \\
& ~ iff ~ & $a \notin \mathbf{outr}_{\mu}(s)$ and $\mu!a, s \models \sigma)$, and \\
& & $a \notin \mathbf{outr}_{\mu}(s)$ and $\mu!a, s \models \sigma')$ \\
& ~ iff ~ & $\mu, s \models \sigma!a$ and $\mu, s \models \sigma'!a$ 
\end{tabular}\end{quote}

The remaining cases are similar. \qed 
\end{proof}

\begin{proof}[Proposition \ref{combined}]
We start with considering the cases where the restrictions are applied consecutively twice. We use Proposition 2.\vspace{-0.1cm}
\begin{quote} \begin{tabular}{lll}
$\mu, s \models (\sigma ! a )!a$ & ~ iff ~ & $a \notin \mathbf{outr}_{\mu}(s) $ and $\mu!a, s \models \sigma ! a$ \\
& ~ iff ~ & $a \notin \mathbf{outr}_{\mu}(s) $ and $(\mathbf{out}_{\mu}(s) \neq a$ and $(\mu!a)!a, s \models \sigma)$ \\ 
& ~ iff ~ & $a \notin \mathbf{outr}_{\mu}(s) $ and $\mu!a, s \models \sigma$ \\
& ~ iff ~ & $\mu, s \models \sigma ! a$
\end{tabular}\end{quote}

Now, we generalize to the case for $n$. The proof is by induction on $n$. The case for $n=1$ is trivial and the case for $n=2$ was presented above. Assume now that 
$n\geq 2$ and the claim holds for every integer less than or equal to $n$. We will now show it for $n+1$.\vspace{-0.1cm}
\begin{quote} \begin{tabular}{llll}
$\mu, s \models \sigma !^{n} a $ & ~ iff ~ &  &\\
$\mu, s \models (\sigma!^{n-1}a)!a$ & ~ iff ~ & $a \notin \mathbf{outr}_{\mu}(s) $ and $\mu!a, s \models \sigma!^{n-1}a$ & \\
& ~ iff ~ &  $a \notin \mathbf{outr}_{\mu}(s) $ and $\mu!a, s \models \sigma!^a$ & \hfill ~ ~ induction hyp. \\
& ~ iff ~ & $\mu, s \models \sigma ! a$ & \hfill by definition
\end{tabular}\end{quote} \qed \end{proof}

\begin{proof}[Proposition \ref{mixed}]
First, notice that for a strategy $\mu$ and moves $a, b$, we have $(\mu!a)!b = (\mu!b)!a$ by definition. Now, let us consider strategy specification $\sigma$ with moves $a$ and $b$.

\begin{quote} \begin{tabular}{lll}
$\mu, s \models (\sigma ! a )!b$ & ~ iff ~ & $b \notin \mathbf{outr}_{\mu}(s)$ and $\mu!b, s \models \sigma ! a$ \\
& ~ iff ~ & $b \notin \mathbf{outr}_{\mu}(s) $ and $(a \notin \mathbf{outr}_{\mu}(s)$ and $(\mu!b)!a, s \models \sigma)$ \\ 
& ~ iff ~ & $b \notin \mathbf{outr}_{\mu}(s) $ and $(a \notin \mathbf{outr}_{\mu}(s)$ and $(\mu!a)!b, s \models \sigma)$ \\ 
& ~ iff ~ & $a \notin \mathbf{outr}_{\mu}(s) $ and $(b \notin \mathbf{outr}_{\mu}(s)$ and $(\mu!a)!b, s \models \sigma)$ \\ 
& ~ iff ~ & $a \notin \mathbf{outr}_{\mu}(s) $ and $\mu!a, s \models \sigma!b$ \\
& ~ iff ~ & $\mu, s \models (\sigma !b)! a$
\end{tabular}\end{quote} \qed \end{proof}

\begin{proof}[Theorem \ref{bgv-completeness}]
The completeness of RSL is by reduction to SL. Since we have one additional strategy specification, we describe an immediate reduction for that. 

First notice that an RSL model is a submodel of a suitable SL model. The only problem is that, in RSL, strategies are constructed as relations whereas in SL, the strategies are functions. Therefore, an RSL strategy can be thought of as a union of several SL strategies. Therefore, a RSL game-tree is a tree model for a SL model with the suitable union of strategies. Moreover, in RSL, we also obtain a game-tree model with relational strategies.

Now, the reduction of RSL to SL should be rather immediate. Given a formula of the form $(\sigma)_i : c$ where $\sigma = \sigma' ! a$ for some $a$, we observe that (by soundness), it is equivalent to a formula in the language of SL: $\mathbf{turn}_i \wedge \neg ((\sigma)_i : a) \wedge (\sigma)_i : c$. Similarly, consider the given formula of the form $\sigma \rightsquigarrow_{i} \psi$ where $\sigma = \sigma' ! a$ for some $a$. Notice also that, in SL,  $\sigma \rightsquigarrow_{i} \psi$ is axiomatically reduced to a formula that uses formulas of the form $(\sigma)_i : a$ which we have covered just before. Therefore, all two different types of formulas that may include restricted strategy specifications are reduced to a formula in the language of SL. Since SL is complete already \cite{gho}, and our translation is truth preserving due to soundness, we conclude that RSL is complete. \qed \end{proof}

\begin{proof}[Theorem \ref{bgv-spec-translation}]
Let $\mu : S \rightarrow \Sigma$ be a strategy and let $\sigma$ be a strategy specification in SL for a fixed player $i$. For simplicity, we omit the superscripts that indicate the player. First, we describe how to obtain a multi-modal CTL* tree $\mu^{*}$, and then show that the translation is truth-preserving. As a reminder, in a strategy tree, we include the root, and for the states that belong to the strategizing player, we assign a unique move to the player. In addition, we include all other moves that do not belong to the strategizing player. Therefore, a strategy tree can be thought of a as branching multi-modal CTL* model. We put $w \models \mathbf{turn}_{i}$ if $w \in S^{i}$ and at each $w \in S^{i}$, we have one outgoing edge. For $v \notin S^{i}$, we have all admissible moves $\mu(v)$ at $v$. Then, the CTL* model is constructed as follows. Take $S$ as the set of states of the CTL* model $\mu^{*}$. Next, for all moves $a \in \Sigma$, we have a corresponding accessibility relation $R_{a}$ in the CTL* model similar to the Kripke semantics for propositional dynamic logic. Finally, we keep the valuation the same as in the SL model. Now, let us consider the case $\sigma = [\psi \rightarrow a]^{i}$:
\begin{quote}\begin{tabular}{lll}
$\mu, s \models \sigma$ & ~ iff ~ & $\mu, s \models [\psi \rightarrow a]^{i}$\\
& ~ iff ~ & $\mu, s \models \psi$ implies $\mathbf{out}_{\mu} = a$ \hfill (by definition) \\
& ~ iff ~ & $\mu^{*}, s \models \mathbf{Tr}(\psi)$ implies $\mathsf{E} (\mathbf{strategy}_{\mu} \wedge \mathsf{X}_{a} \top)$\hfill
\end{tabular} \end{quote}
Here we make use of the formula translation $\mathbf{Tr}$ whose correctness will be shown next. Furthermore, we obtain the last line by the induction hypothesis, and by the earlier observation we have made about $\mathbf{out}_{\mu} = a$. The remaining cases for the translation $\mathbf{tr}$ are straightforward inductions on specifications $\sigma_{1} + \sigma_{2}$ and $\sigma_{1} \cdot \sigma_{2}$, and hence left to the reader. Unconventionally, here we make use of Theorem~\ref{bgv-translation} in the proof of the above statement just because specifications preceed the formulas in SL.  \qed \end{proof}

\begin{proof}[Theorem \ref{bgv-translation}]
Take a SL model $M$ and the corresponding multi-modal CTL* model $M^{*}$. We then have the following:
\begin{quote}\begin{tabular}{lllr}
$M, s \models \langle a \rangle \varphi$ & iff & $\exists s'$ s.t. $s \stackrel{a}{\Rightarrow} s'$ and $M, s' \models \varphi$ & \hfill (by definition) \\
& & $\exists s'$ s.t. $s \stackrel{a}{\Rightarrow} s'$ and $M^{*}, s' \models \mathbf{Tr}(\varphi)$ & \hfill (by induction) \\
& & $M^{*}, s \models \mathsf{X}_{a} \mathbf{Tr}(\varphi)$ & \hfill (by definition)\\
\end{tabular}\end{quote}
Similarly, consider the SL formula $\sigma \rightsquigarrow_{i} \psi$:
\begin{quote}\begin{tabular}{lll}
$M, s \models \sigma \rightsquigarrow_{i} \psi$ & iff & for all $s'$ such that $s \Rightarrow^{*}_{\sigma} s'$ in $T_s | \sigma$, we have,\\
&& $M, s' \models \psi \wedge (\mathbf{turn}_i \rightarrow \mathbf{enabled}_{\sigma})$ \\ && (by definition) \\
&iff & for all $s'$ such that $s \Rightarrow^{*} s'$ in $T_{s}$, we have \\
&& $M, s' \models \mathbf{strategy}_{\sigma} \wedge \psi \wedge (\mathbf{turn}_i \rightarrow \mathbf{enabled}_{\sigma})$ \\
&& (by definition) \\
&iff & for all $s'$ such that $s \Rightarrow^{*} s'$ in $T_{s}$, we have \\
&& $M^{*}, s' \models \mathbf{strategy}_{\sigma} \wedge \mathbf{Tr} (\psi) \wedge (\mathbf{turn}_i \rightarrow \mathbf{enabled}_{\sigma})$\\
&& (by induction)\\
&iff & $M^{*}, s \models \mathsf{E} \Box [\mathbf{strategy}_{\sigma} \wedge \mathbf{Tr}(\psi) \wedge (\mathbf{turn}_i \rightarrow \mathbf{enabled}_{\sigma})]$ \\
&& (as $\Rightarrow^{*}$ denotes the reflexive and transitive closure \\
&& on the path $T_{s}$ on the CTL* tree $T$) \vspace{-0.1cm}\\
\end{tabular}\end{quote}

The last case $(\sigma)_i : c$ is very similar. Recall that this formula is true in SL at a state $s$ iff $c \in \sigma(s)$. Consider the enabled move $c$ at $s$, and take a corresponding next time modality for $c$, namely $\mathsf{X}_{c}$. Moreover, at $s$, the diamond-like modality $\mathsf{X}_{c}$ has to be \emph{enabled} giving us $\mathsf{X}_{c} \top$ for $c \in \sigma(s)$. \qed \end{proof}

\begin{proof}[Theorem \ref{bgv-decide}]
In this proof, we will present two different arguments to show that the model checking problem for SL is in PSPACE.

First, we show that the translation $\mathbf{Tr}$: SL $\rightarrow$ multi-modal CTL* is polynomial-time in terms of the length of the formulas. We will denote the length of a specification or a formula by $|\cdot|$. We will show that $|\mathbf{Tr}(\psi)| \leq |\psi|^k$ for some integer $k$. The cases for Booleans are obvious, hence we skip them. For $\psi = \langle a \rangle \varphi$, consider $|\mathbf{Tr}(\langle a \rangle \varphi)|$ which is equivalent to $|\mathsf{X}_a \mathbf{Tr}(\varphi)| = 1 + |\mathbf{Tr}(\varphi)|$. By induction hypothesis, $|\mathbf{Tr}(\varphi)| \leq |\varphi|^l$ for some integer $l$. Therefore, $1 + |\mathbf{Tr}(\varphi)| \leq |\varphi|^{k}$ for some $k\geq 1+l$. Therefore, for $\psi = \langle a \rangle \varphi$, we observe $|\mathbf{Tr}(\psi)| \leq |\psi|^k$ for some integer $k$. In a similar fashion, the case for $\psi = (\sigma)_i : c$ is obvious as $|\mathbf{Tr}((\sigma)_i : c)|$ is always constant. The case for $\psi = \sigma \rightsquigarrow_i \varphi$ is also very similar. By induction hypothesis, we immediately observe that $|\mathbf{Tr}(\sigma \rightsquigarrow_i \varphi)| \leq 11 + |\sigma \rightsquigarrow_i \varphi|^l$ (counting the parantheses as well). Therefore, for some large enough integer $k > l$, we have $|\mathbf{Tr}(\sigma \rightsquigarrow_i \varphi)| \leq |\sigma \rightsquigarrow_i \varphi|^k$.

Now, we can consider the translation $\mathbf{tr}$ for the strategy specifications. The reason why we consider $\mathbf{tr}$ after $\mathbf{Tr}$ is the fact that former depends on the latter. We will now show that, for strategy specification $\sigma$, $|\mathbf{tr}(\sigma)| \leq |\sigma|^k$ for some integer $k$. Consider the case where $\sigma = [\psi \rightarrow a]^i$. Then, $|\mathbf{tr}([\psi \rightarrow a]^i)| = 5 + |\mathbf{Tr}(\psi)|$. By the previous observation, we know that $|\mathbf{Tr}(\psi)| \leq |\psi|^l$ for some integer $l$. Therefore, $5 + |\mathbf{Tr}(\psi)| \leq |\psi|^k$ for large enough integer $k \geq l$. Thus, $\mathbf{tr}(\psi \rightarrow a]^i) \leq |(\psi \rightarrow a]^i)|^k$ for some $k$. The cases for the $\cdot$ and $+$ operations are obvious. This concludes the proof that the translation functions $\mathbf{tr}$ and $\mathbf{Tr}$ are polynomial-time. Therefore, the complexity of model checking for SL cannot be higher than PSPACE.

Our second argument is a sketch of a direct reasoning. Similar to the arguments for the complexity of basic modal logic, we just need to check the branches of the tree model one by one using a depth-first search \cite{bla1}. Now, fix a SL formula $\varphi$ and a model $M$. Since SL models are branching tree models, we can check $M$, branch by branch, one at a time without any need of considering the other branches. Namely, the procedure does not need to \emph{remember} the previous searches making it effective in the use of space. Moreover, the length of the branches in $M$ is polynomial in $|\varphi|$ because of the construction of the SL model $M$, thus it shows that model checking for SL is in PSPACE. This concludes the proof that the complexity of the model checking problem for SL is in PSPACE. \qed
\end{proof}

\begin{proof}[Corollary \ref{bgv-exponential}]
We observed that the RSL formulas can be reduced to SL formulas. The reduction is clearly polynomial, as can be seen from the axiomatization (See Section~\ref{RSL-axiom}). Therefore, we can translate any given RSL formula to a SL in a SL model, which in turn can be translated into a multi-modal CTL* formula and model. Then, by Theorem~\ref{bgv-decide}, we deduce that the complexity of the model checking problem for RSL is also PSPACE. \qed
\end{proof}
\end{document}